\documentclass[runningheads]{llncs}
\usepackage{graphicx}
\usepackage{hyperref}

\begin{document}
\title{Look Ma, no code: fine tuning nnU-Net for the AutoPET II challenge by only adjusting its JSON plans} 
\titlerunning{Fine tuning nnU-Net for the AutoPET II challenge}
%
\author{Fabian Isensee\inst{1,2} \and
Klaus H. Maier-Hein\inst{1,2,3}}
\authorrunning{Isensee et al.}
%
\institute{Division of Medical Image Computing, German Cancer Research Center (DKFZ), Heidelberg, Germany \and
Helmholtz Imaging, German Cancer Research Center (DKFZ), Heidelberg, Germany \and
Pattern Analysis and Learning Group, Department of Radiation Oncology, Heidelberg University Hospital, Heidelberg, Germany}
\maketitle              
\begin{abstract}
We participate in the AutoPET II challenge by modifying nnU-Net only through its easy to understand and modify 'nnUNetPlans.json' file. By switching to a UNet with residual encoder, increasing the batch size and increasing the patch size we obtain a configuration that substantially outperforms the automatically configured nnU-Net baseline (5-fold cross-validation Dice score of 65.14 vs 33.28) at the expense of increased compute requirements for model training. Our final submission ensembles the two most promising configurations.

\end{abstract}
\section{Introduction}
The AutoPET II challenge provides a large training dataset for segmenting tumor lesions located all across the body. Input modalities are CT and PET scans. There are 1014 training cases available stemming from 900 patients. The test set comprises 200 images, 50 of which are drawn from the same distribution as the train set while the remaining 150 are taken from other data sources. There is also a preliminary test set consisting of 5 cases. Besides the raw images, the organizers provide preprocessed images pairs where the PET and CT scans are coregistered and cropped to the same geometry and where the PET scan intensity values are transformed into standard update values (SUVs). For more information please refer to the \href{https://autopet-ii.grand-challenge.org/}{challenge homepage}.

nnU-Net \cite{isensee2021nnu} is a framework that automates the configuration of UNet \cite{ronneberger2015u} based segmentation pipelines for a given dataset. It has seen tremendous success in the past, being used regularly to compete in and win segmentation competitions \cite{zhao2021coarse,andrearczyk2021overview,full2021studying,isensee2019attempt,isensee2023extending,isensee2021nnubrats}. It was originally developed for the Medical Segmentation Decathlon \cite{antonelli2022medical} where it demonstrated exceptional generalization across datasets.

In addition to its fully automated configuration capabilities, nnU-Net supports easy interaction with its generated configurations via JSON files which allows users to fine tune it further without having to touch any source code. As our contribution to the AutoPET II we set out to explore how far we take this features. Specifically, we constrain our model search space to modifying the nnU-Net generated 'nnUNetPlans.json' file. No changes to the source code are made.

\section{Methods}
We use the provided preprocessed images and convert them into the nnU-Net data format. The 'modality' for both CT and PET/SUV images it set to 'CT' in order to tell nnU-Net that intensities of both inputs represent standardized physical quantities. We then run the default nnU-Net planning and preprocessing steps. For more details on nnU-Net, see the original nnU-Net paper \cite{isensee2021nnu} or read the documentation in the \href{https://github.com/MIC-DKFZ/nnUNet}{GitHub repository}.

\subsection{Residual Encoder UNet}
Ever since our participation in KiTS2019 \cite{heller2021state,isensee2019attempt} nnU-Net has build in support for U-Nets with residual encoders. We investigate whether the default Residual Encoder UNet generated by nnU-Net outperforms its plain UNet architecture. 

\subsection{Scaling with batch size}
We experiment with increasing batch size, from the default of 2 all the way to 80. Thanks to nnU-Net's ability to let configurations inherit from each other, each new configuration only required three lines of intuitive text changes.

\subsection{Scaling the patch size}
Another axis along which nnU-Net can be scaled to more compute is the patch size. We increase the patch size from the automatically configured 128x128x128 to 192x192x192 voxels. Even though no changes to the network topology are made, the default networks configured by nnU-Net have a sufficiently large receptive field to make effective use of the larger inputs. 

\section{Results}

\begin{figure}[t]
    \includegraphics[width=\textwidth]{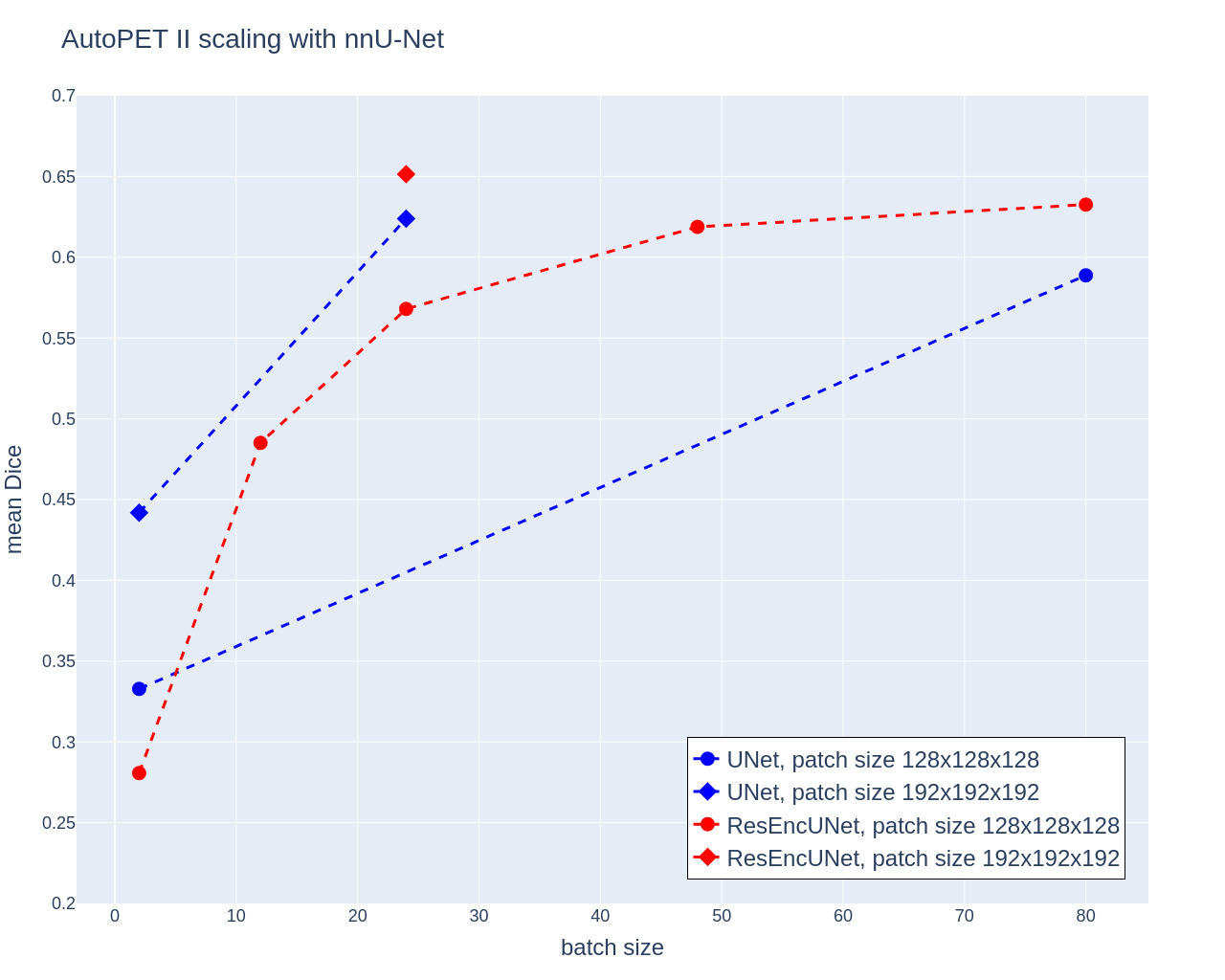}
    \caption{The Residual Encoder UNet consistently outperforms a UNet with standard convolutional encoder. Increasing the batch size and the patch size both yield substantial improvements over the respective baseline values. Increasing patch size is more effective than the batch size. Best results are achieved with scaling both. All modifications to nnU-Net are made by editing the 'nnUNetPlans.json file'. No code changes were required.}
    \label{fig1}
\end{figure}

All experiments are run via the standard nnU-Net 5-fold cross-validation on the 1014 training cases (stratified by patients). Models are trained from scratch and no external data was used for training. Aside from the aforementioned modifications, no changes to the nnU-Net defaults are made. All changes are accessible through modifying the 'nnUNetPlans.json' file and require no code modifications.

Figure \ref{fig1} summarizes our cross-validation experiments. We refer to it throughout the remainder of this section.

\subsection{Residual Encoder UNet}
With the default batch size of 2, the Residual Encoder UNet (red circle, Dice 28.07) appears to be outperformed by the default plain UNet (blue circle, Dice 33.28). However, upon closer inspection, fold 4 of the Residual Encoder UNet failed to converge. On the remaining four folds the Residual Encoder UNet outperformed the plain UNet (Dice 35.01 vs 33.53). This failure to converge is not observed in any of the other experiments. Across all batch and patch size scaling experiments the Residual Encoder UNet consistently outperforms its plain UNet counterpart.

\subsection{Scaling with batch size}
As we can see from the batch size scaling experiments (see red line with circles as markers) there is a clear advantage of scaling compute via increasing the batch size. The Dice score improves from 28.07 all the way to 63.26 for the Residual Encoder UNet. We observe diminishing returns with increasing batch size: from batch size 2 to 12 the Dice score increases from 28.07 to 48.51 whereas from 48 to 80 it merely increases from 61.88 to 63.26.

\subsection{Scaling the patch size}
For the regular UNet and batch size 2, switching to 192x192x192 sized inputs improved the mean Dice score from 33.28 to 44.19. In our experiments, networks with larger patch size consistently outperform those with smaller inputs.

\subsection{Putting it together}
Our results indicate that a residual encoder is superior to a plain UNet on the AutoPET II dataset. Moreover we see that increasing both the batch and patch size in isolation yield substantial improvements over the baseline. We therefore train the Residual Encoder UNet with the largest batch size our GPU hardware can manage for each of the explored patch sizes. This yields a configuration trained with patch size 128x128x128 and batch size 80 (Dice 63.26) and a configuration with patch size 192x192x192 and a batch size of 24 (Dice 65.14). Training one model took about a week on 8xNvidia A100 40GB GPUs. 

For our final submission we ensemble these models. Like all nnU-Net configurations, these models were trained via 5-fold cross-validation on the training cases rather than one model on all data. Thus, the total number of models in our ensemble is 10.

To cut down on inference time, we use a step size of 0.6 for the models with patch size 128x128x128. This parameter controls how much the sliding window is shifted between predictions as a function of the patch size. We also restrict test time data augmentation in the form of mirroring to the sagittal and coronal axes. Since the preliminary test set consists of just five cases we only used it to verify that our Docker image completes the inference within the 10 minutes time limit per case imposed by the challenge organizers. On this preliminary test set our algorithm ranks 7th with a Dice score of 56.50, false negative volume 0.0398 and false positive volume 0.0000. Note that the official evaluation awards correctly predicted empty segmentation masks (no false positives) a Dice score of 0 whereas nnU-Net would exclude those cases from mean aggregation.

\section{Discussion}
We build our solution using nnU-Net \cite{isensee2021nnu}, a powerful tool for automating the design of U-Net based segmentation pipelines and potent framework for method development. By design we have constricted the development space of our solution to simple changes in the 'nnUNetPlans.json' file. Not a single line of code was written to build our method. By switching to a Residual Encoder UNet, increasing the batch size and increasing the patch size, we were able to substantially improve upon the automatically configured nnU-Net baseline at the expense of increased compute requirements for model training. Our final submission consists of an ensemble of the two best nnU-Net configurations totaling 10 individual models.

What constitutes a good model in the context of a competition is defined by the metrics and ranking schemes used to select the winners. However, due to a lack of time, the official evaluation scheme was not used for model selection in this work. All decisions were simply made based on the mean Dice score of the cross-validation, omitting the false positive and false negative predicted volumes. Furthermore, we have not used the rank-then-aggregate scheme to compare algorithms, although prior work has shown that this could be beneficial \cite{isensee2021nnubrats}. The disregard for the official evaluation scheme is not good practice and could have resulted in sub-par decision making. 

By design we have constricted modifications to nnU-Net to its simple to understand and modify JSON files. Our intention was to demonstrate what nnU-Net can do without touching its code. Naturally, without the code editing constraint, a plethora of additional modifications to nnU-Net's data augmentation, sampling strategy and loss function could have unlocked higher segmentation accuracy at a lower compute budget. 

The domain transfer aspect of the challenge is not covered in our experiments. For participants without a lot of experience in PET imaging, sample images of how such shifts could affect the images would have been useful. Even images from data sources different than those of the test images would have been informative and could have steered method development. In our method, we leave it to the predefined data augmentation implemented in nnU-Net to make the models as robust as possible. Future work could investigate whether the findings from \cite{full2021studying} (i.e. batch normalization + strong data augmentation) benefit the domain robustness in CT+PET lesion segmentation. 

While scaling up batch and patch size proved to be an effective strategy on AutoPET II, scaling on other datasets has so far produced less pronounced gains. Particularly considering the high compute requirements relative to the standard nnU-Net (which only requires a single GPU with 10GB of VRAM), adapting nnU-Net to use more resources by default would yield minimal gains on average while greatly reducing the potential audience for nnU-Net. We believe that providing simple access to crucial hyperparameters via JSON files strikes a good balance and empowers researchers to adjust them when necessary. 

Pretrained models as well as a detailed description for how the 'nnUNetPlans.json' file was adapted are provided via the nnU-Net documentation  \href{https://github.com/MIC-DKFZ/nnUNet/blob/master/documentation/competitions/AutoPETII.md}{nnU-Net documentation}.

\section*{Acknowledgment} 
Part of this work was funded by Helmholtz Imaging, a platform of the Helmholtz Incubator on Information and Data Science.

\bibliographystyle{splncs04}
\bibliography{mybibliography}

\end{document}